\documentclass[prb,twocolumn, superscriptaddress,nobibnotes]{revtex4-1}
\usepackage{graphicx,amsmath,amssymb,amsxtra,times,subfigure,color}
\renewcommand{\narrowtext}{\begin{multicols}{2} \global\columnwidth20.5pc}

\let\textquotedbl="

\begin{document}

\title{Demagnetization Borne Microscale Skyrmions}

\author{Patrick Johnson}

\affiliation{Department of Physics, Washington University, St. Louis, MO 63130,
USA}

\author{A. K. Gangopadhyay}

\affiliation{Department of Physics and Center for Materials Innovation, Washington University, St. Louis, MO 63130,
USA}

\author{Ramki Kalyanaraman}

\affiliation{Department of Materials Science and Engineering,
University of Tennessee, Knoxville, TN 37996, USA}

\affiliation{Department of Chemical and Biomolecular Engineering,
University of Tennessee, Knoxville, TN 37996}

\author{Zohar Nussinov}
\email[E-mail at: ]{zohar@wuphys.wustl.edu}
\affiliation{Department of Physics and Center for Materials Innovation, Washington University, St. Louis, MO 63130,
USA}

\date{\today}
\begin{abstract}
Magnetic systems are an exciting realm of study that is being explored
on smaller and smaller scales. One extremely interesting magnetic
state that has gained momentum in recent years is the skyrmionic state.
It is characterized by a vortex where the edge magnetic moments point
opposite to the core. Although skyrmions have many possible realizations, in
practice, creating them in a lab is a difficult task to accomplish.
In this work, new methods for skyrmion generation and customization
are suggested. Skyrmionic behavior was numerically observed in minimally 
customized simulations of spheres, hemisphere, ellipsoids, and hemi-ellipsoids, for typical Cobalt 
parameters, in a range from  approximately $40 \: nm$ to $120 \: nm$ in 
diameter simply by applying a field. 
\end{abstract}
\maketitle
\section[]{Introduction}
\label{Introduction}
A skyrmion, theorized first by Skyrme in 1962  \cite{THR1962556}, is a state with a vectorial order parameter 
which is aligned at the system boundary at an opposite direction to what the order parameter assumes 
at the origin. Skyrmions may appear in diverse arenas, such as elementary particles
\cite{THR1962556,Atiyah1989438,Houghton1998507,PhysRevLett.79.363,2002hep.ph....2250W}, 
liquid crystals \cite{RevModPhys.61.385}, Bose-Einstein condensates \cite{Khawaja2001,PhysRevA.62.013602,PhysRevA.68.043602}, thin magnetic 
films \cite{0022-3727-44-39-392001}, quantum Hall systems 
\cite{PhysRevB.47.16419,PhysRevLett.75.2562,PhysRevLett.75.4290,PhysRevLett.74.5112}, 
and potentially vortex lattices in type II superconductors \cite{RevModPhys.76.975,2011arXiv1108.3562B}. 
Being able to experimentally observe or generate skyrmions is a current research thrust 
\cite{THR1962556,THR1962556,Atiyah1989438,Houghton1998507,PhysRevLett.79.363,2002hep.ph....2250W,RevModPhys.61.385,Khawaja2001,PhysRevA.62.013602,PhysRevA.68.043602,0022-3727-44-39-392001,PhysRevB.47.16419,PhysRevLett.75.2562,PhysRevLett.75.4290,PhysRevLett.74.5112,RevModPhys.76.975,2011arXiv1108.3562B,PhysRevLett.103.250401,Schulz2012,Kirakosyan2006413}. 

In this work we demonstrate via micromagnetic simulations that achieving a 
skyrmion is as simple as creating a nanoparticle of many possible
geometries, which is large enough to support a single vortex but small enough 
to prevent multiple vortices. The demagnetization energy allows for
the formation of a vortex at zero-field.  We find that as the field increases such that it lies
in a direction opposite to the core, the magnetization at the edges may realign itself parallel to the field
direction more readily than the magnetization next to the core. Immediately prior to annihilation of the vortex 
(i.e., the flipping of the magnetization at the system core to become parallel to the applied
field direction), the skyrmionic state is most notable. We observed this, relatively ubiquitous, effect in systems with disparate geometries-
spheres, hemispheres, ellipsoids, and hemi-ellipsoids.  It may be possible to generalize 
this process so as to experimentally synthesize a skyrmion lattice by simply creating 
an array of nanoparticles with tunable size and spacing, such as by 
self-organzation \cite{krishna:073902,Krishna2011356}. Preliminary simulations 
of a two-by-two grid of Cobalt hemispheres of radius $20 \: nm$  with varying inter-hemisphere
separation indicate that beyond a threshold distance of twice the radius, an array of 
skyrmions is formed.  As the center to center separation is steadily increased, the skyrmionic state becomes more
 lucid. For small separations, interactions partially thwart the creation of the individual skyrmions.

As is well known, we can quantify a skyrmionic state by calculating the Pontryagin
index (also known as a winding number) that is given by \cite{eduardo1999field}

\begin{eqnarray}
Q=\frac{1}{8\pi}\int d^{2}x\epsilon_{ij}\hat{M}\cdot(\partial_{i}\hat{M}\times\partial_{j}\hat{M}).
\label{PontryaginIndex}
\end{eqnarray}

In this expression, $\epsilon_{ij}$ is the two dimensional anti-symmetric tensor and $\hat{M}$
is the normalized magnetization.  For a single skyrmion,
this winding number (or topological charge) is equal to unity. Skyrmions are characterized
by the non-trivial homotopy class $\pi_{2}(S^{2})$. This homotopy
class is characterized by an integer that, for this case, is the
Pontryagin index. States with different integer skyrmion number (the
Pontryagin index) cannot be continuously deformed into one another.

In the current context, the skyrmionic state resides on a two dimensional plane. On each spatial point of the plane, 
there is a three dimensional order parameter which, in our case, is the magnetization
$\vec{M}$. Topologically, a skyrmion is a magnetic state such
that when it is mapped onto a sphere (via stereographic
projection) resembles a monopole or hairy ball.
This means that on mapping from a flat space to the surface of a sphere,
the individual magnetic moments will always point perpendicular to
the surface of the sphere, much like a magnetic monopole.

The above topological classification is valid for an ``ideal'' skyrmion on an infinite two-dimensional plane or disk  with the condition that
the local moment $\vec{M}(\vec{r})$ at spatial infinity (irrespective of the location $\vec{r}$ on the infinite disk) all orient in the same direction:
$\lim_{r\to\infty}\hat{M}({\vec{r}})=\hat{M}_{0}$. In such a case $\hat{M}_{0}$ corresponds to the magnetization at the 
``point at infinity''.  On applying a stereographic
projection of the infinite plane onto a unit sphere, $\hat{M}_{0}$
maps onto the magnetization at the north pole of the unit sphere while the oppositely oriented 
$\hat{M}$ at the origin corresponds to the magnetization at the south pole. In such a case, 
the winding number is identically equal (in absolute value) to unity. In many physically pertinent geometries, including
the systems simulated in this work, there are finite size limits which
only allow the magnetization $\vec{M}$ to exhibit the trend of approaching a uniform value $\vec{M}_{0}$
as one moves away from the center of the system. In this case, the integral in Eq. 
\ref{PontryaginIndex} is not an integer.  However,  it is clear that, in the limit of 
infinite planar size, these states would become ideal skyrmions and the winding 
number $Q$ would approach an integer value.

The remainder of this article is organized as follows. In Section \ref{Theory}, we 
provide necessary background. We briefly describe the simulations employed 
in this work and discuss energetic considerations.  Section \ref{ResultsandDiscussion} reports on our 
central result- the numerical observation of skyrmions. We discuss a higher 
dimensional generalization and the possibility of generating skyrmion lattices. 
We conclude in section \ref{Conclusion} with a summary of our results.

\section[]{Theory}
\label{Theory}
\subsection[]{Simulation Theory} 
\label{Simulation Theory}
In this work of simulating magnetic states
of nanoparticles, the Object Oriented Micromagnetic Framework (OOMMF)
1.2a distribution as provided from NIST was utilized \cite{Donahue1999}.
The OOMMF code numerically solves the Landau-Lifshitz Ordinary Differential
Equation given by,

\begin{eqnarray}
\frac{d\vec{M}}{dt}=-|\bar{\gamma}|\vec{M}\times\vec{H}_{eff}-\frac{|\bar{\gamma}|\tilde{\alpha}}{M_{s}}\vec{M}\times\left(\vec{M}\times\vec{H}_{eff}\right)
\end{eqnarray}
 where $\vec{M}$ is the magnetization, $\bar{\gamma}$ is the Landau-Lifshitz
gyromagnetic ratio, $M_{s}$ is the saturation magnetization, $\tilde{\alpha}$
is the damping coefficient, and $H_{eff}$ is the effective field
given by derivatives of the Gibbs free energy. The Gibbs free energy,
in this case, is given by \cite{Brown1978},

\begin{eqnarray}
G=\int(\frac{1}{2}C\left[\left(\vec{\nabla}\alpha\right)^{2}+\left(\vec{\nabla}\beta\right)^{2}+\left(\vec{\nabla}\gamma\right)^{2}\right]+w_{a}\nonumber \\
-\frac{1}{2}\vec{M}\cdot\vec{H}'-\vec{M}\cdot\vec{H}_{0})d\tau
\end{eqnarray}
 where $\alpha$, $\beta$, and $\gamma$ are the directional cosines,
$C$ is proportional to the exchange stiffness constant and depends
on the crystal structure, $w_{a}$ is the crystalline anisotropy term,
$\vec{H}'$ is the demagnetization field, and $\vec{H}_{0}$ is the
external magnetic field. The crystalline anisotropy term can be expressed
in terms of anisotropy constants, $K_{1}$ and $K_{2}$, and directional
cosines as,

\begin{eqnarray}
w_{a}=K_{1}\left(\alpha^{2}\beta^{2}+\beta^{2}\gamma^{2}+\gamma^{2}\alpha^{2}\right)+K_{2}\alpha^{2}\beta^{2}\gamma^{2}.
\end{eqnarray}

In the simulations, a metastable state was determined to have been
reached when the maximum torque experienced by any one magnetic moment,
measured in $\frac{degrees}{ns}$, dropped below $0.2$. Once this
level of torque was reached, the magnetic state data were saved to
a file along with the other properties of the system, including but not limited
to, the energies associated with each contribution, overall magnetization,
and number of iterations. The magnetic field was then changed to the next value 
and the iterations continued until saturation of the magnetization was obtained.  The
magnetic field steps were chosen such that
half the steps (typically, a few hundred) were during the increasing field portion and the other
half in the decreasing field portion. The data stored in the file were used later to generate
the hysteresis plots, track the energy changes associated with the field variations, and the 
spatial orientations of the magnetic moments.  Unless specified otherwise, the parameters 
chosen in the simulations correspond to those for Cobalt,
as shown in Table \ref{CobaltParameters}.

\begin{table}
\begin{tabular}{|p{3.5cm}|p{2cm}|p{2cm}|}
\hline 
parameter   & value used in this work\tabularnewline
\hline 
Exchange Stiffness Constant ($A$)  & $2.5\times10^{-11} \frac{J}{m}$   \tabularnewline
Saturation Magnetization ($M_{s}$)  & $1.4\times10^{6} \frac{A}{m}$   \tabularnewline
Crystalline Anisotropy Constant ($K_{1}$)  & $5.20\times10^{5} \frac{J}{m^{3}}$   \tabularnewline
Damping Constant ($\tilde{\alpha}$)  & $0.5$ \tabularnewline
Landau-Lifshitz Gyromagnetic Ratio ($\bar{\gamma}$)   & $2.21\times10^{5} \frac{m}{A\cdot s}$  \tabularnewline
Stopping Torque ($\frac{dm}{dt}$)  & $0.19 \frac{deg}{ns}$   \tabularnewline
\hline 
\end{tabular}\caption{Table of parameters used in the simulations of particles in this work.
The exchange stiffness constant, saturation magnetization, and crystalline
anisotropy constant are material specific and are chosen for Cobalt.
The damping constant, Landau-Lifshitz-Giblert gyromagnetic ratio,
and stopping torque are material independent parameters.}

\label{CobaltParameters} 
\end{table}

\subsection[]{Energy Considerations} 
\label{Energy Considerations}
In our simulations, we considered field, demagnetization, and exchange
energies. For simplicity, we neglected crystalline anisotropy effects.
The field tries
to align the local magnetic moments parallel to it while exchange effects
favor an alignment of the magnetic moments with
their nearest neighbors. The (universally geometry borne) demagnetization energy
directly relates to dipole-dipole interactions
\cite{Brown1978}. Demagnetization energy is often the dominant
term for long range behaviors while exchange effects tend to dominate
at short spatial scales.

As is well known, the competition between the long range and the short range energy contributions leads to
the creation of domain walls. The demagnetization favors
oppositely oriented moments at the expense of exchange effects that favor
slow variations amongst neighbors. Ultimately, this tradeoff gives rise
to domain walls in micromagnetic systems.

The potential energy from demagnetization of a system is given by

\begin{eqnarray}
\mathcal{E}_{M}=-\frac{1}{2}\sum_{i}\vec{m}_{i}\cdot\vec{h}'_{i},
\end{eqnarray}
where $\vec{h}'_{i}$ is the effective field at position $i$ that originates from all other
dipoles. This field can be written as

\begin{eqnarray}
\vec{h}'_{i}=\vec{H}'+\frac{4}{3}\pi\vec{M}+\vec{h}''_{i},
\end{eqnarray}
 where $\vec{H}'$ is the megascopic field from the poles due to $\vec{M}$
outside of a physically small sphere around site $i$. The second
term subtracts the effective field inside an arbitrary small region (or sphere) centered about point $i$, 
and $\vec{h}''_{i}$ is the field at site $i$ created by dipoles
inside this region. In general, $\vec{h}''_{i}$ depends
on the crystal lattice structure. In the continuum limit, the sum
becomes an integral of the form,

\begin{eqnarray}
\mathcal{E}_{M}=-\frac{1}{2}\int\vec{M}\cdot(\vec{H}'+\frac{4}{3}\pi\vec{M}+\Lambda\cdot\vec{M})dV.
\end{eqnarray}

The tensor $\Lambda$ in the third term depends only on the crystal
structure and local magnetization and can grouped with crystalline anisotropy.
This tensor also vanishes for cubic crystals identically.
The second term in this expression is a constant proportional to $M_{s}^{2}$ and can be ignored. 
The $\Lambda$ tensor also vanishes for cubic crystals identically leaving,

\begin{eqnarray}
\mathcal{E}_{M}=-\frac{1}{2}\int\vec{M}\cdot\vec{H}'dV
\label{EnergyEquation}.
\end{eqnarray}

The demagnetization field, $\vec{H}'$, can equivalently be derived from Maxwell's equations.  It can be expressed 
as the negative gradient of a potential, $U$ that satisfies the equations,

\begin{eqnarray}
\nabla^{2}U_{in}=\gamma_{B}\vec{\nabla}\cdot\vec{M}\\
\nabla^{2}U_{out}=0,
\end{eqnarray}
 with the surface boundary conditions,

\begin{eqnarray}
U_{in}=U_{out}\\
\frac{\partial U_{in}}{\partial n}-\frac{\partial U_{out}}{\partial n}=\gamma_{B}\vec{M}\cdot\vec{n}.
\end{eqnarray}
where the constant $\gamma_{B}$ is, in our units, $4 \pi$.

 Lastly, the potential needs to be regular at infinity, such that
$|rU|$ and $|r^{2}U|$ are bounded as $r\rightarrow\infty$.
Our simulations directly capture the demagnetization field effects. 

From the standpoint of energy, for a skyrmion to be possible, the
dimensions of the ellipsoid must be larger than the critical dimensions
at which vortices can nucleate in a given system.  For example, for the
hemispherical geometry, with the typical values of Table
\ref{CobaltParameters}, the critical radius
was found to be $19 \: nm$. For larger radii,
vortices are the preferred state before reaching zero field. The vortex
will nucleate such that the core is parallel to the field and the
remainder of the vortex lies in the plane perpendicular to the field. Once
the field begins to oppose the direction of the moments
at the core, the energy cost of eliminating the core is significantly
higher than allowing the outer magnetic moments to align more with
the field. When the exchange energy cost of the skyrmionic state becomes
greater than the demagnetization energy for a uniform magnetization,
the core flips, annihilating the skyrmion, and the magnetization saturates.
Immediately, prior to this, though, a skyrmionic state can be achieved.

Ezawa \cite{PhysRevLett.105.197202} raised the specter of a skyrmionic
state in thin films via the computation of the energy of such assumed
variational states within a field theoretic framework of a non-linear
sigma model. Dipole-dipole interactions may stabilize such a state
below a critical field. Our exact numerical calculations for the evolution
of the magnetic states demonstrate that not only are skyrmionic states
viable structures, but are actually the precise lowest energy state
for slices of hemispheres and other general structures.

\section[]{Results and Discussion}
\label{ResultsandDiscussion}
\subsection[]{Observation of a Skyrmion} 
\label{Observation of a Skyrmion}
\begin{figure}[htbp]
 \centering \includegraphics[scale=0.26]{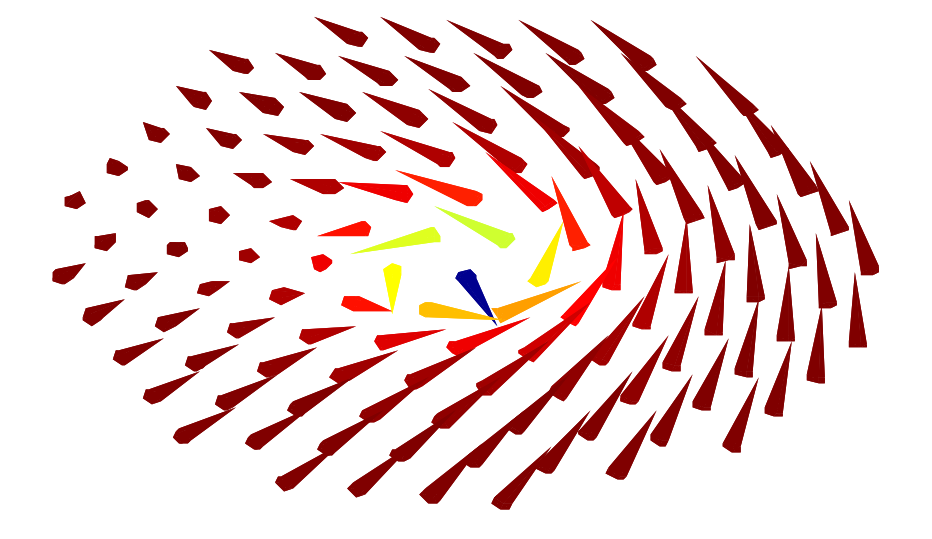}
\caption[]{Vector plot of the skyrmion state for the bottom slice of a hemisphere
of radius $24 \: nm$. Not all local magnetic moments are shown for the sake
of clarity.}

\label{Hemisphere24nmSkyrmionBottomLayerNoAxes} 
\end{figure}

As our numerical simulations vividly illustrate, just prior to the annihilation of the vortex, 
the magnetic moments at the edge of the system start to orient themselves in a direction 
opposite to that in the core. On increasing the radius of the simulated hemispheres and 
spheres, the configurations next to the basal plane better conformed to the full skyrmion 
topology (i.e., that on an infinite plane).t should be noted here, that
as the radius of a hemisphere increases, the crossover to a double
vortex state will eventually occur, but if one vortex is maintained,
in the limit of large radii, a full skyrmion would be expected. 
This may be possible in materials with large exchange constant and 
small saturation magnetization.  In what follows, we will employ the typical values appearing in Table \ref{CobaltParameters}.
The skyrmion state for the bottom layer (basal plane) of a hemisphere of radius
$24 \: nm$ is shown in Figure \ref{Hemisphere24nmSkyrmionBottomLayerNoAxes}.

\begin{figure}[htbp]
 \centering \includegraphics[scale=0.25]{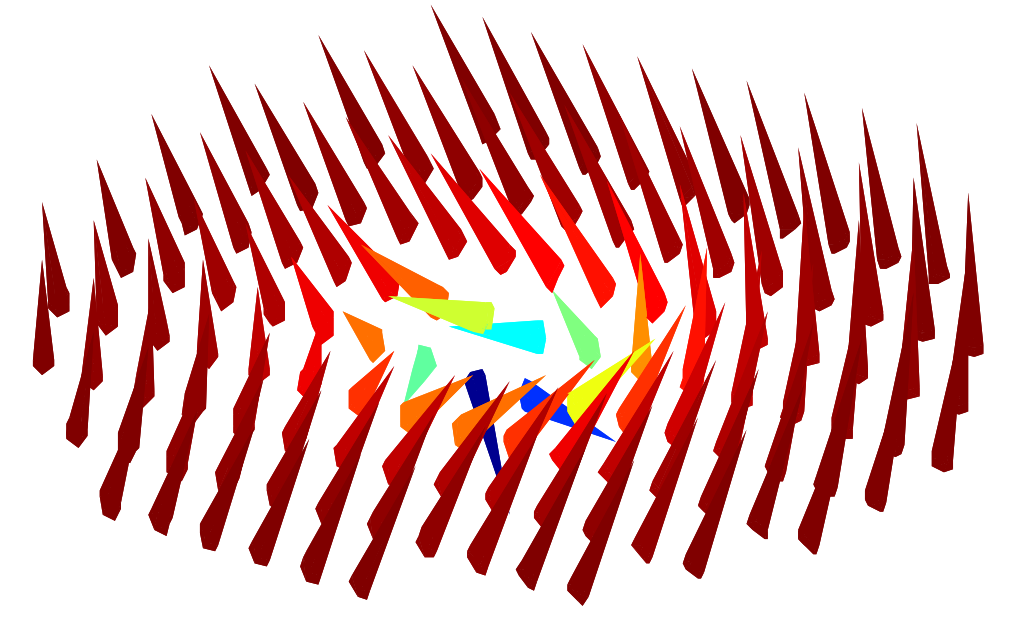}
\caption[]{Vector plot of the skyrmion state in a sphere of radius $59nm$.
The slice is along the equator of the sphere. Only a subset of local
magnetic moments is shown for clarity.}

\label{SphereSkyrmion} 
\end{figure}

A similar configuration was observed in simulation runs for nanospheres. For a sphere,
symmetry does not favor any particular direction, but that symmetry
is broken once a field is applied. Skyrmions were observed in runs
of spheres large enough to support a vortex which corresponds to a
radius of $\approx15 \: nm$. As the radius of the sphere increases, 
the edge magnetic moments and the core magnetic
moments become more antiparallel. A skyrmion in a sphere of radius $59nm$ is shown
in Figure \ref{SphereSkyrmion}.

\begin{figure}[htpb!]
 \centering \includegraphics[scale=0.19]{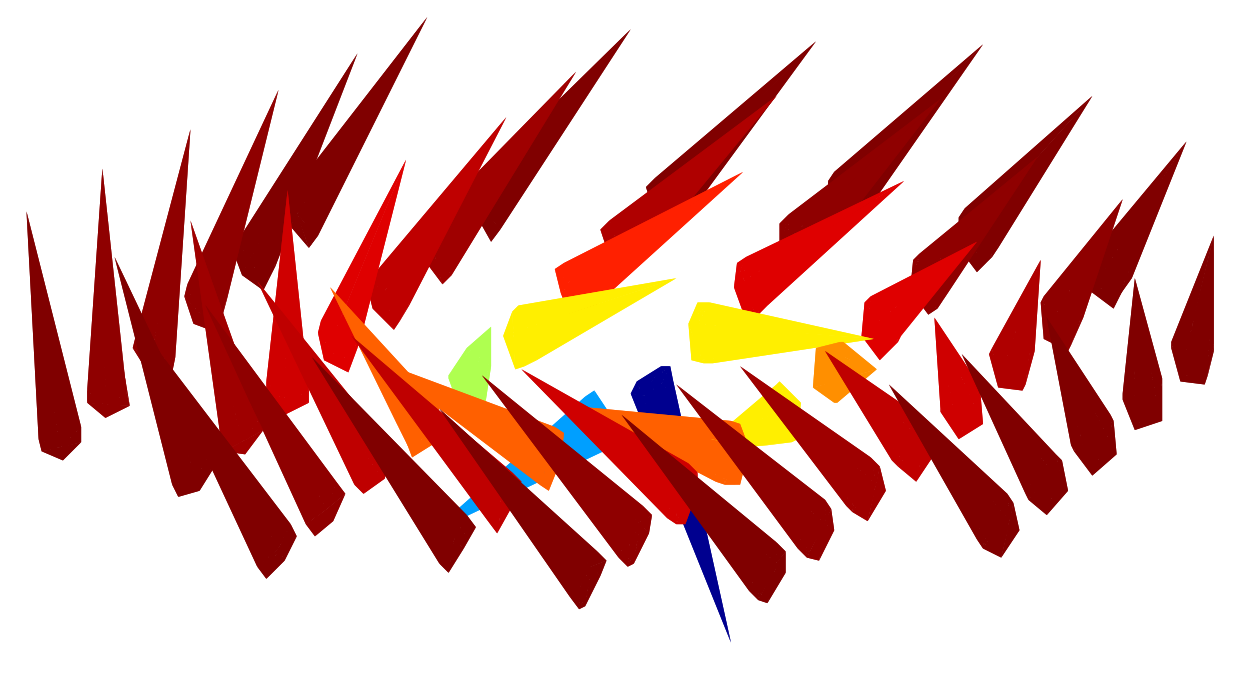}
\caption{Vector plot of the skyrmion state in an ellipsoid with major axis
of $20 \: nm$ and minor axis of $15 \: nm$. The slice is along the equator
of the ellipsoid. Only a subset of local magnetic moments is shown
for clarity.}

\label{EllipsoidSkyrmion} 
\end{figure}

\begin{figure}[htpb!]
 \centering \includegraphics[scale=0.23]{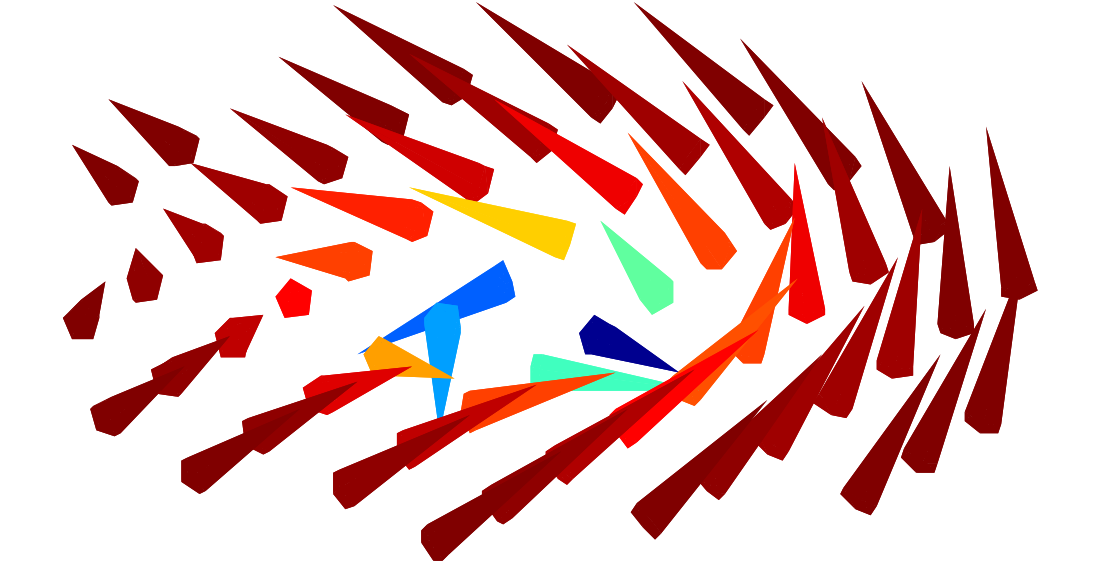}
\caption{Vector plot of the skyrmion state in a hemi-ellipsoid with major axis
of $20nm$ and minor axis of $15 \: nm$. The slice is along the base
of the hemi-ellipsoid. Only a subset of local magnetic moments is
shown for clarity.}

\label{HemiellipsoidSkyrmion} 
\end{figure}

Once skyrmions were observed in these systems, it
begged the question, ``Do these occur in ellipsoids and
hemi-ellipsoids?" Upon examining this, indeed skyrmions
can be observed in oblate ellipsoids and hemi-ellipsoids as shown in
Figures \ref{EllipsoidSkyrmion} and \ref{HemiellipsoidSkyrmion}.

\begin{figure}[htbp]
 \centering \includegraphics[scale=0.35]{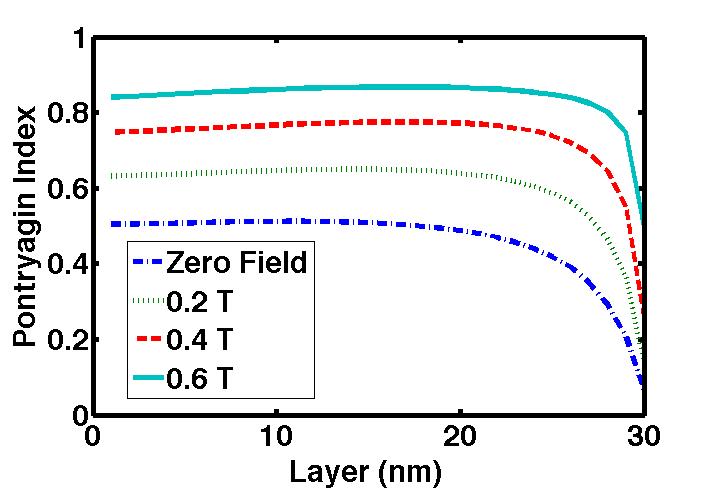}
\caption{Plot of the Pontryagin index versus the z-coordinate of the slice
taken from the hemisphere of radius $30 \: nm$. These are shown for increasing field from
zero field (dark blue dot-dash line), $0.2 \: T$ (green dotted line), $0.4 \: T$ (red dashed line), and $0.6 \: T$
(teal solid line).}

\label{PontryaginPlot} 
\end{figure}

To verify that these are structures approach those of skyrmions and to quantitively 
monitor their deviations from an ideal skyrmionic state (for which the Pontryfin index 
is unity),we computed the Pontryagin index at different cross sections of the hemisphere. 
These cross sections were those of the hemisphere with planes parallel to the basal 
plane(i.e., that at the base of the hemisphere). For a hemisphere with radius $30 \: nm$, 
we calculated the skyrmion number Q for thirty individual parallel layers vertically 
separated by $1 \: nm$. We numerically evaluated the integral of Eq. \ref{PontryaginIndex} 
for all of these layers and examined how it changes as the field increases from $0$ to $0.8\: T$.
These data are shown in Figure \ref{PontryaginPlot}.

\begin{figure*}[htbp]
\centering 
\subfigure[]{ 
\includegraphics[scale=0.29]{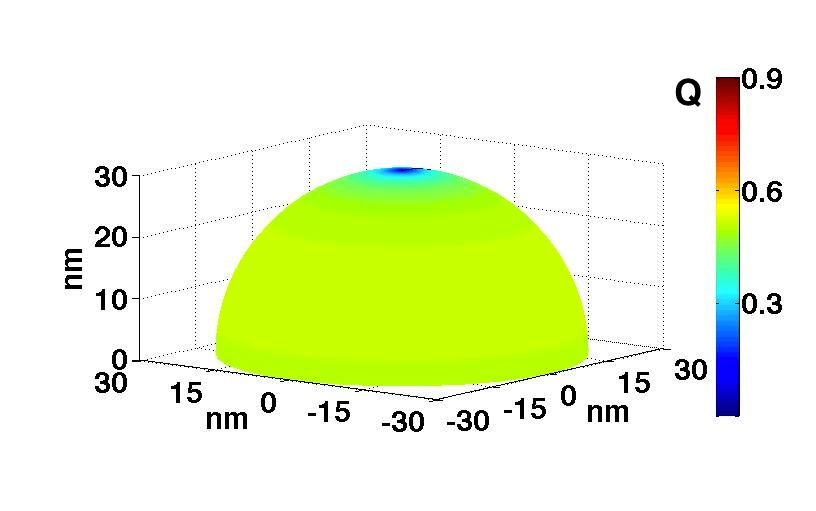}
\label{PontryaginHemisphere50} } 
\subfigure[]{ 
\includegraphics[scale=0.31]{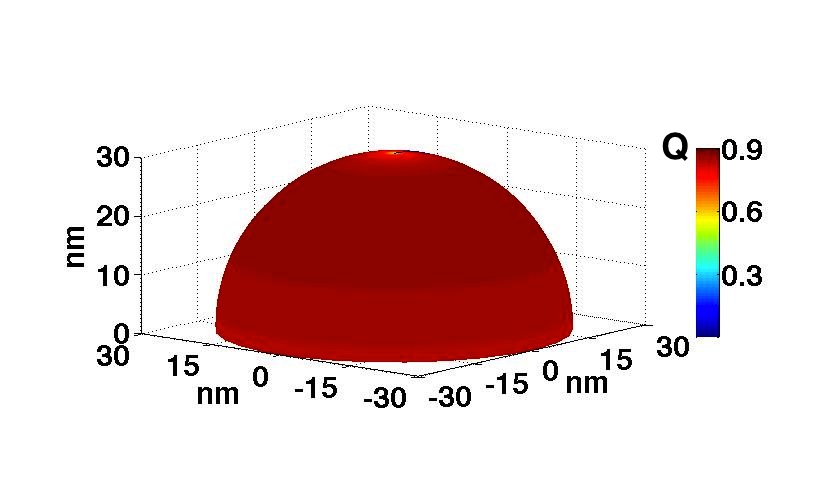}
\label{PontryaginHemisphere80} } 
\caption{Three dimensional plots of the Pontryagin index for a hemisphere of radius 
$30 \: nm$ at (a) zero field and (b) $0.6 \: T$}

\label{PontryaginHemisphere30nm} 
\end{figure*}

Visualizing this in the geometry of the hemisphere specifically, one
can look at how the Pontryagin index varies along various planes of a hemisphere,
starting from the equator and moving to the pole. 
It can be clearly seen that the skyrmionic behavior exists
for most of the height of the hemisphere and only the cap deviates
from the rest of the system. The size of this cap depends on the given
field strength as can be seen in the case of 0 field (Figure \ref{PontryaginHemisphere50})
and with a field of $0.6 \: T$ (Figure \ref{PontryaginHemisphere80}).  At higher fields, 
prior to the annihilation of the vortex, the Pontryagin index approaches an integer value, 
as expected for an ideal skyrmionic state.

\begin{figure*}[htbp]
\centering 
\subfigure[]{ 
\includegraphics[scale=0.3]{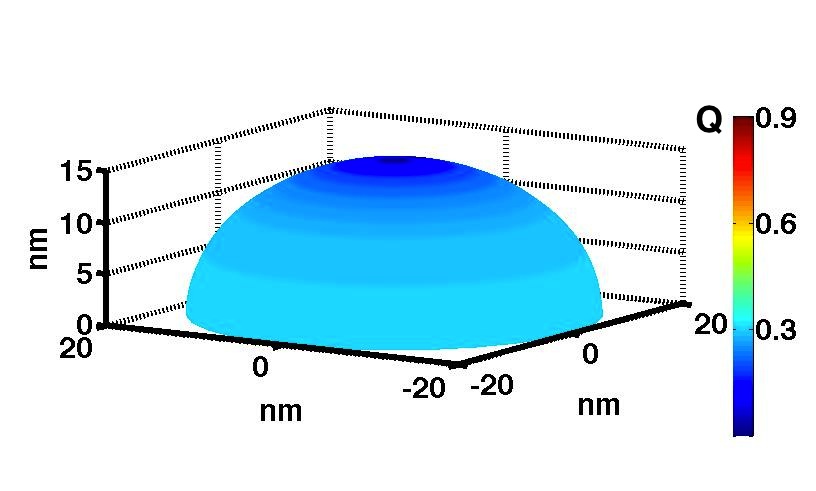}
\label{PontryaginHemiellipsoid20nm15nm} } 
\subfigure[]{ 
\includegraphics[scale=0.29]{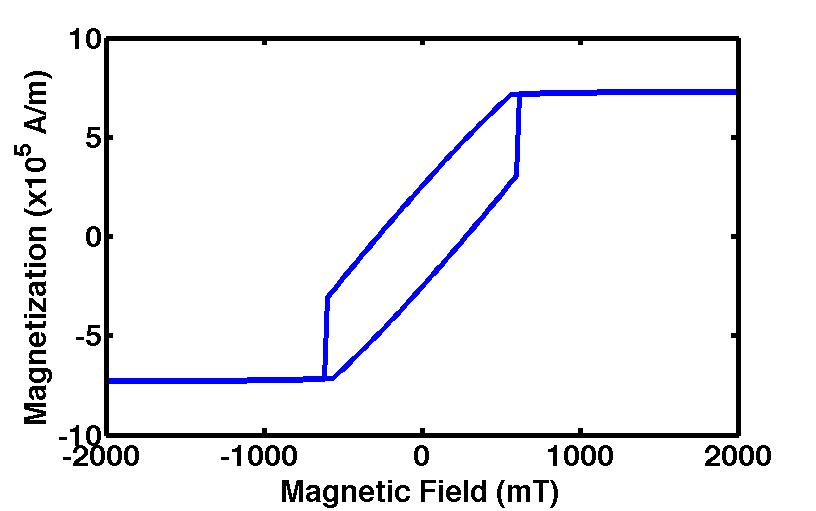}
\label{Hemiellipsoid20nm15nmHysteresis} } 
\subfigure[]{ 
\includegraphics[scale=0.29]{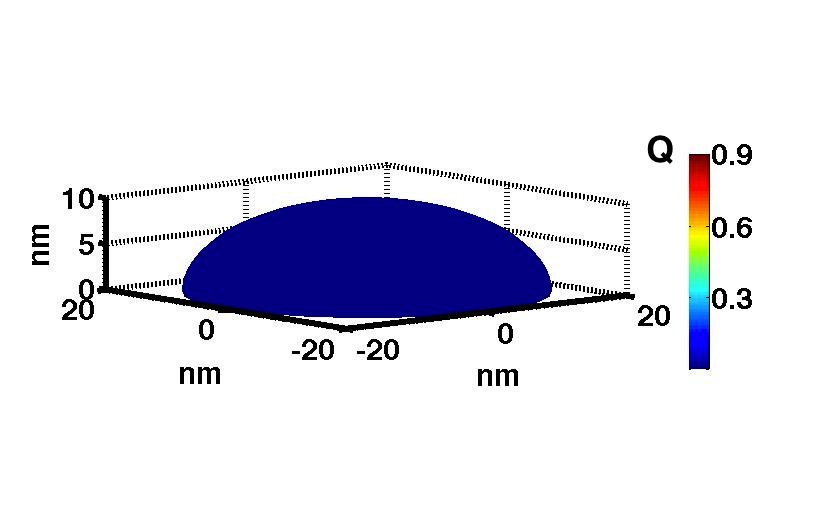}
\label{PontryaginHemiellipsoid20nm10nm} } 
\subfigure[]{ 
\includegraphics[scale=0.29]{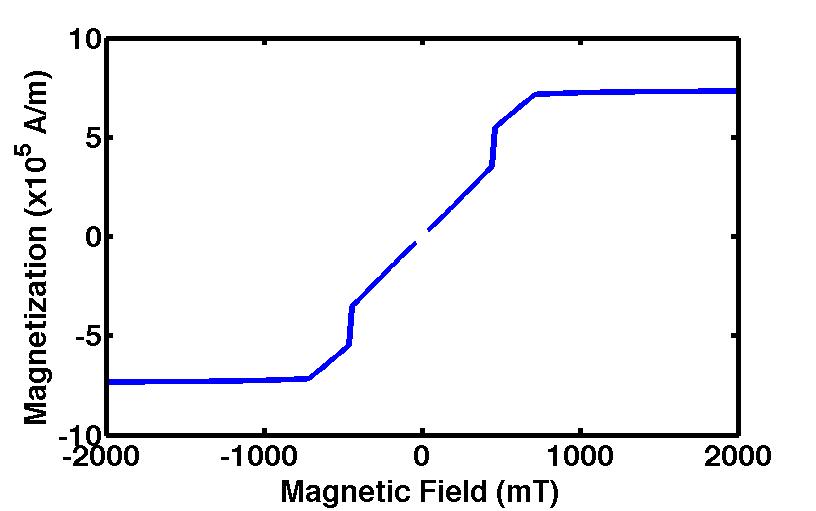}
\label{Hemiellipsoid20nm10nmHysteresis} } 
\subfigure[]{ 
\includegraphics[scale=0.31]{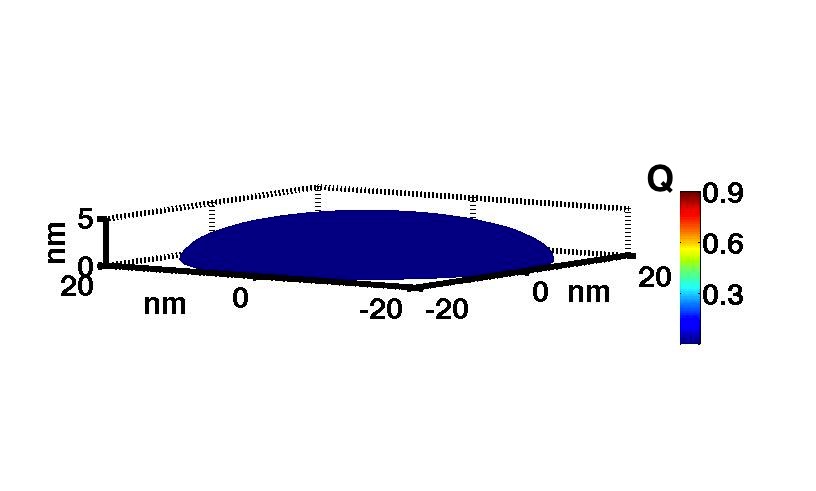}
\label{PontryaginHemiellipsoid20nm5nm} } 
\subfigure[]{ 
\includegraphics[scale=0.29]{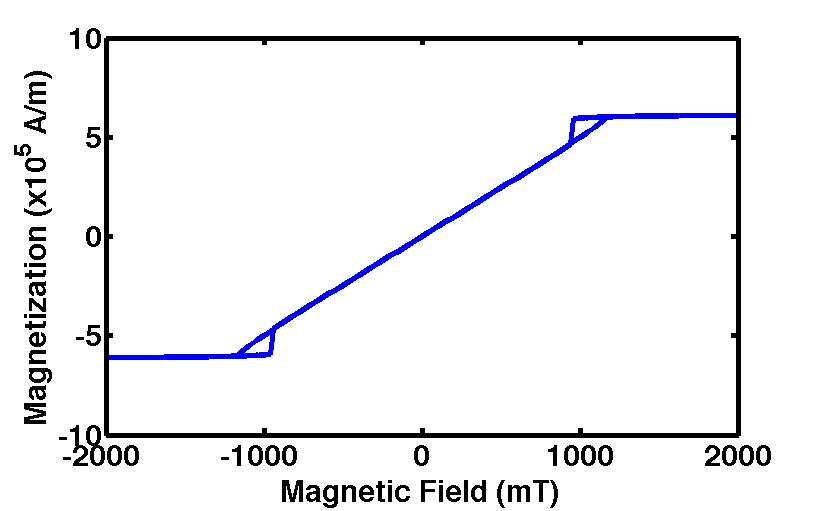}
\label{Hemiellipsoid20nm5nmHysteresis} 
} 
\caption{Plots of the Pontryagin index and how it varies with height inside 
hemi-ellipsoids of $30 \: nm$ radius major axis as the minor axis varies from $15 \: nm$ (a) to $10 \: nm$
(c) to $5 \: nm$ (e). This is shown for (a) field equal to $0.2 \: T$ pointing in the negative z-direction 
(perpendicular to the face of the hemi-ellipsoids). As will be noted, the
existence of skyrmionic behavior is not prevalent in the more flattened
hemiellipsoinds and vanishes at this field between minor axis $15 \: nm$ and $10 \: nm$.  
The associated partial hysteresis loops for each of these hemi-ellipsoid runs are shown in Figs. (b), (d), 
and (f), respectively.}
\label{PontryaginHemiellipsoid} 
\end{figure*}

Performing similar analysis on the hemi-ellipsoids and visualizing
the Pontryagin index and its variance with height, it can be seen
that the same behavior exists in a less extreme way than the hemispheres.
This behavior can be seen in Figure \ref{PontryaginHemiellipsoid} for hemi-ellipsoids
of fixed $30 \: nm$ major axis and varying minor axis.

In examining the hysteresis behavior of the hemi-ellipsoids, one can
see a trend as the z-dimension goes from the hemisphere radius
($20 \: nm$) to the minimum simulated size of $5 \: nm$. This trend shows
a movement from extensive vortex and skyrmionic behavior in the more
hemispheric geometries and less vortex and skyrmionic behavior in the more
ellipsoidal geometries.

Although it will not  be considered in this work, crystalline anisotropy could influence 
the formation of skyrmions in a number of ways.  In the case of a single 
crystal, the vortex state would be more difficult to nucleate and thus the
skyrmionic state is less energetically favorable. When many 
crystalline grains are present, the results discussed here are valid as the large number
or randomly oriented crystals will, on average, not favor any direction,
and thus will not favor any one direction.

\subsection[]{Generalization to a Hedgehog} 
\label{Generalization to a Hedgehog}
These results lead to the question
of whether this can be generalized to more than two dimensions. The
natural generalization from the two-dimensional skyrmion to a three-dimensional
magnetic state would be the hedgehog. The hedgehog resides in three spatial dimensions
coupled with a three dimensional order parameter.
The canonical example of a hedgehog is$\vec{M}=M_{s}\hat{r}$
where the magnetization always points outwards. A skyrmion is related
to a hedgehog via a stereographic projection from the sphere onto
a plane where the south pole of the hedgehog projects to the core
of the skyrmion on the plane and the north pole of the hedgehog projects
to the points at infinity on the plane. Calculating the demagnetizing
field for this state in a sphere gives rise to a potential and field
equal to

\begin{eqnarray}
U(r)=\gamma_{B}M_{s}(r-R),\\
\vec{H}=-\gamma_{B}M_{s}\hat{r}.
\end{eqnarray}

Plugging this into Equation \ref{EnergyEquation}, one finds the energy
of the hedgehog to be $2\pi M_{s}^{2}(4\pi/3)R^{3}$. Comparing this
to the energy of the uniformly magnetized state, $(1/2)(4\pi/3)^{2}M_{s}^{2}R^{3}$,
it can easily be seen that the hedgehog has three times the energy
of the uniform state. This, combined with the fact that the exchange
energy and the field energy will favor the uniform state, the hedgehog
state will not be possible in a sphere.

If one were to continuously deform the hedgehog by rotating the local
magnetic moments by $\pi/2$ such that $\vec{M}=M_{s}f(z)\hat{\phi}$
where $f(z)$ is a function that goes to 0 as $z\rightarrow0$ such
that the exchange energy does not diverge, one would find the demagnetization
energy of that state to be identically 0. The field energy in this
system is also 0 for a field that is applied along the z-axis. The
exchange energy is given by $(4\pi/3)RC$ where $C$ is the exchange
stiffness constant. The total energy of this state is equal to the
exchange energy, and comparing this to the uniform state, a hedgehog
of this form is favorable for,

\begin{eqnarray}
R\ge\sqrt{\frac{C}{\frac{2\pi\mu_{0}M^{2}}{3}-MH_{0}}}.
\end{eqnarray}

For $C=2.5\times10^{-11}J/m$ and $M_{s}=1.4\times10^{6}A/m$ as it
is for Cobalt, at $0$ field, this radius works out to be $\approx3.5\mu m$.

\subsection[]{Skyrmion Array} 
\label{Skyrmion Array}
It is illuminating to consider the possibility of an array of skyrmions.  
As briefly discussed below, we find that effective particle interactions 
may thwart the creation of a skyrmion lattice when these particles are 
not far separated. However, for sufficiently large center to center separations, a Skryme 
lattice may be achieved.  In preliminary simulations of arrays of 
nanoparticle arrays, simulations of a two-by-two grid of hemispheres of 
radius $20 \: nm$ with a variable separation show that a center to center separation of four times 
the radius is close enough that the nanoparticles still interact magnetically 
and prevent the formation of an array of skyrmions. As expected, further 
separation should approach the the single particle result of skyrmions, 
as we briefly discuss next. 

\begin{figure*}[htbp]
\centering 
\subfigure[]{ S
\includegraphics[scale=0.25]{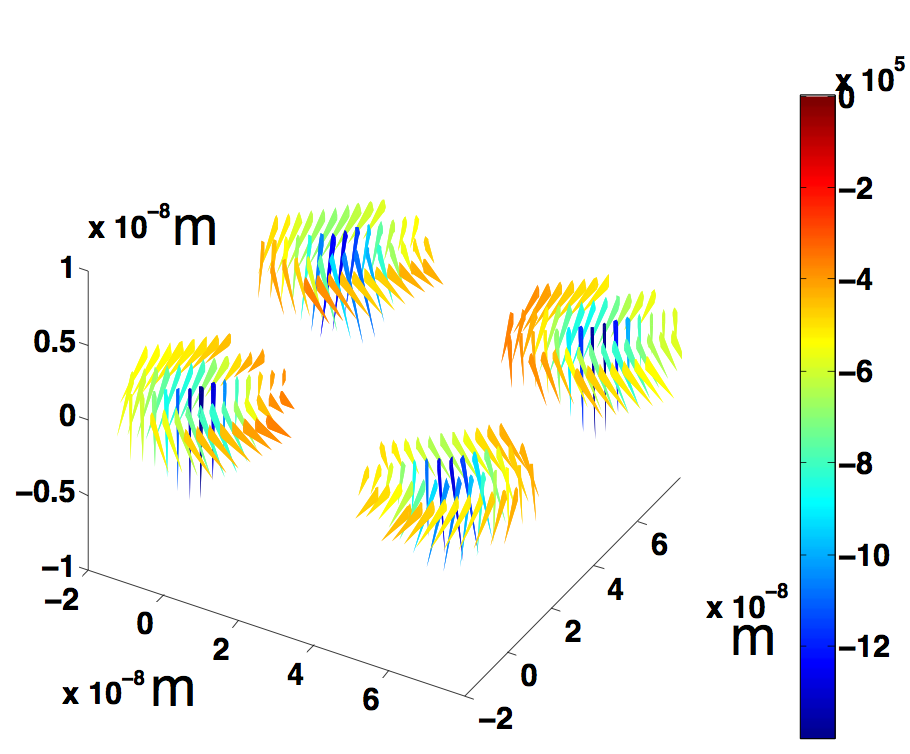}
\label{4Hemisphere25nmField94} 
} 
\subfigure[]{ 
\includegraphics[scale=0.25]{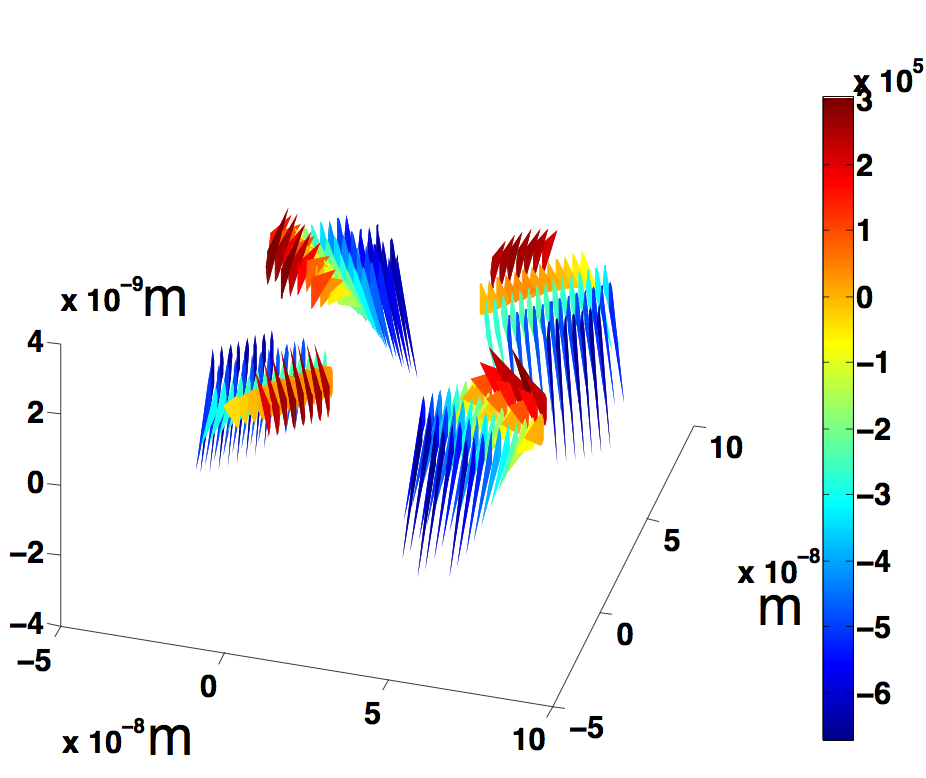}
\label{4Hemisphere25nmField95} 
} 
\caption{Vector plot of a 2x2 array of hemispheres with radius $20 \: nm$ and center to center separation $80 \: nm$ at fields of $0.12 \: T$ pointing in the negative z-direction (a) and $0.1\: T$ pointing in the negative z-direction (b).  Colorscale corresponds to z-component of the local magnetic moment in units of $A/m$.}
\label{4ArrayFields94And95} 
\end{figure*}

The transition from the array of particles which support individual
vortices to the array of particles that are clearly interacting with
each other can be seen in Figure \ref{4ArrayFields94And95}. In this
figure, the annihilation of the vortices can be seen as the particles
realign their magnetization to form a state where the local magnetization
orients in the counterclockwise direction from particle to particle,
yet within each particle, when moving in the counterclockwise direction,
the local magnetization changes from oriented in the negative z-direction
to the positive z-direction.

\begin{figure*}[htbp]
\centering 
\subfigure[]{ 
\includegraphics[scale=0.6]{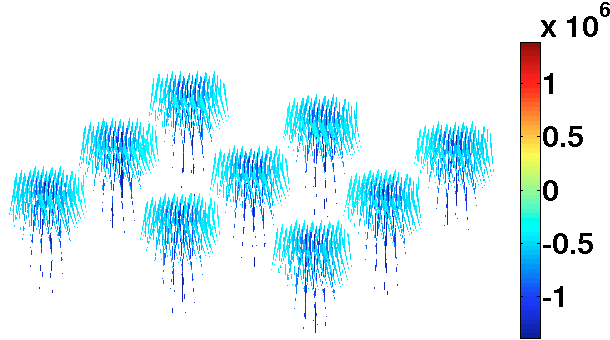}
\label{Grid9HemisphereField95} 
} 
\subfigure[]{ 
\includegraphics[scale=0.6]{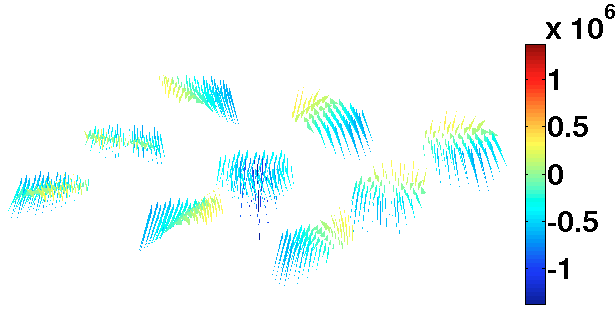}
\label{Grid9HemisphereField96} 
} 
\caption{Vector plot of a 3x3 array of hemispheres with radius $20 \: nm$ and center to center separation $80 \: nm$ at fields of $0.1 \: T$ pointing in the negative z-direction (a) and $0.08\: T$ pointing in the negative z-direction (b).  Colorscale corresponds to z-component of the local magnetic moment in units of $A/m$.}
\label{9ArrayFields95And96} 
\end{figure*}

In repeating these simulations for a 3x3 array of hemispherical nanoparticles,
the same behavior was observed.  This array was similar to the 2x2 array in
that it had nanoparticles with diameters of $40 \: nm$ and center to center
separation of $80 \: nm$.  The annihilation of the vortices occurred at 
a slightly larger field (0.08T rather than 0.1T) as shown in Fig \ref{9ArrayFields95And96}.

\section[]{Conclusion}
\label{Conclusion}

We conclude with a brief synopsis of our findings. We carried a systematic 
numerical study of the magnetization of small nanoparticles in the presence 
of an external magnetic field. These systems were simulated for different 
sizes and geometry (sphere, hemisphere, ellipsoids). Our analysis 
ignored anisotropy (crystalline, shape, strain, etc.) effects. We find that, 
as has been widely reported in the literature 
\cite{Shinjo11082000,hubert1998magnetic}, beyond a critical diameter, 
the particles enter into a single vortex state under zero external field; 
multiple vortices are possible for much larger particles.  
Our key new result concerns {\it the creation of skyrmions 
in the single vortex state}. As the field is increased, vortex annihilation 
is accompanied by the formation of a skyrmionic state wherein the 
magnetization of the vortex core points to a direction opposite to that 
at the edge of the nanoparticle. Our result illustrates how geometry plays 
a  pivotal role. Spheres and hemispheres more readily achieve 
skyrmionic states than higher eccentricity ellipsoids. Our preliminary results 
suggested that for center to center separations larger than twice the particle diameters, an 
array of skyrmions may be realized. More detailed studies of skyrmion lattices 
for such particle arrays are planned for the future.

{\bf Acknowledgements.}
Work at Washington University was partially supported by NSF grants 
DMR-1106293 and DMR-0856707, and by the Center for Materials Innovation 
(CMI) of Washington University. Work at the university of Tennessee was partially 
supported by NSF DMR-0856707.

\bibliography{ScalingDraftBib}

\end{document}